\documentclass[aps,prd,preprint,superscriptaddress,tightenlines,nofootinbib]{revtex4}

\usepackage{graphicx}
\usepackage{dcolumn}
\usepackage{bm}

\newcommand{\mevcc}{\!\mathrm{MeV}\!/c^2}

\newcommand{\mev}{\!\mathrm{MeV}}

\newcommand{\gevc}{\!\mathrm{GeV}/\!c}
\newcommand{\gev}{\!\mathrm{GeV}}

\newcommand{\uones}{\Upsilon(1S)}
\newcommand{\utwos}{\Upsilon(2S)}
\newcommand{\uthrees}{\Upsilon(3S)}
\newcommand{\ufours}{\Upsilon(4S)}
\newcommand{\etabones}{\eta_b(1S)}
\newcommand{\etab}{\eta_b}
\newcommand{\buthrees}{\mathcal{B}(\uthrees\to\gamma\etab)}
\newcommand{\butwos}{\mathcal{B}(\utwos\to\gamma\etab)}

\newcommand{\bgammaetab}{\mathcal{B}(\gamma\etab)}
\newcommand{\egammaetab}{E_\gamma(\gamma\etab)}

\newcommand{\calL}{\mathcal{L}}

\newcommand{\Egamma}{E_\gamma}
\newcommand{\Egammaz}{E_0}

\newlength{\figwidth}
\begin{document}

\preprint{CLNS 09/2060}       
\preprint{CLEO 09-13}         

\title{\boldmath Measurement of the $\eta_b(1S)$ mass and the branching fraction for  $\Upsilon(3S)\to\gamma\eta_b(1S)$}

\author{G.~Bonvicini}
\author{D.~Cinabro}
\author{A.~Lincoln}
\author{M.~J.~Smith}
\author{P.~Zhou}
\author{J.~Zhu}
\affiliation{Wayne State University, Detroit, Michigan 48202, USA}
\author{P.~Naik}
\author{J.~Rademacker}
\affiliation{University of Bristol, Bristol BS8 1TL, UK}
\author{D.~M.~Asner}
\author{K.~W.~Edwards}
\author{J.~Reed}
\author{A.~N.~Robichaud}
\author{G.~Tatishvili}
\author{E.~J.~White}
\affiliation{Carleton University, Ottawa, Ontario, Canada K1S 5B6}
\author{R.~A.~Briere}
\author{H.~Vogel}
\affiliation{Carnegie Mellon University, Pittsburgh, Pennsylvania 15213, USA}
\author{P.~U.~E.~Onyisi}
\author{J.~L.~Rosner}
\affiliation{University of Chicago, Chicago, Illinois 60637, USA}
\author{J.~P.~Alexander}
\author{D.~G.~Cassel}
\author{R.~Ehrlich}
\author{L.~Fields}
\author{R.~S.~Galik}
\author{L.~Gibbons}
\author{S.~W.~Gray}
\author{D.~L.~Hartill}
\author{B.~K.~Heltsley}
\author{J.~M.~Hunt}
\author{D.~L.~Kreinick}
\author{V.~E.~Kuznetsov}
\author{J.~Ledoux}
\author{H.~Mahlke-Kr\"uger}
\author{J.~R.~Patterson}
\author{D.~Peterson}
\author{D.~Riley}
\author{A.~Ryd}
\author{A.~J.~Sadoff}
\author{X.~Shi}
\author{S.~Stroiney}
\author{W.~M.~Sun}
\affiliation{Cornell University, Ithaca, New York 14853, USA}
\author{J.~Yelton}
\affiliation{University of Florida, Gainesville, Florida 32611, USA}
\author{P.~Rubin}
\affiliation{George Mason University, Fairfax, Virginia 22030, USA}
\author{N.~Lowrey}
\author{S.~Mehrabyan}
\author{M.~Selen}
\author{J.~Wiss}
\affiliation{University of Illinois, Urbana-Champaign, Illinois 61801, USA}
\author{M.~Kornicer}
\author{R.~E.~Mitchell}
\author{M.~R.~Shepherd}
\author{C.~M.~Tarbert}
\affiliation{Indiana University, Bloomington, Indiana 47405, USA }
\author{D.~Besson}
\affiliation{University of Kansas, Lawrence, Kansas 66045, USA}
\author{T.~K.~Pedlar}
\author{J.~Xavier}
\affiliation{Luther College, Decorah, Iowa 52101, USA}
\author{D.~Cronin-Hennessy}
\author{K.~Y.~Gao}
\author{J.~Hietala}
\author{R.~Poling}
\author{P.~Zweber}
\affiliation{University of Minnesota, Minneapolis, Minnesota 55455, USA}
\author{S.~Dobbs}
\author{Z.~Metreveli}
\author{K.~K.~Seth}
\author{B.~J.~Y.~Tan}
\author{A.~Tomaradze}
\affiliation{Northwestern University, Evanston, Illinois 60208, USA}
\author{S.~Brisbane}
\author{J.~Libby}
\author{L.~Martin}
\author{A.~Powell}
\author{P.~Spradlin}
\author{C.~Thomas}
\author{G.~Wilkinson}
\affiliation{University of Oxford, Oxford OX1 3RH, UK}
\author{H.~Mendez}
\affiliation{University of Puerto Rico, Mayaguez, Puerto Rico 00681}
\author{J.~Y.~Ge}
\author{D.~H.~Miller}
\author{I.~P.~J.~Shipsey}
\author{B.~Xin}
\affiliation{Purdue University, West Lafayette, Indiana 47907, USA}
\author{G.~S.~Adams}
\author{D.~Hu}
\author{B.~Moziak}
\author{J.~Napolitano}
\affiliation{Rensselaer Polytechnic Institute, Troy, New York 12180, USA}
\author{K.~M.~Ecklund}
\affiliation{Rice University, Houston, Texas 77005, USA}
\author{J.~Insler}
\author{H.~Muramatsu}
\author{C.~S.~Park}
\author{E.~H.~Thorndike}
\author{F.~Yang}
\affiliation{University of Rochester, Rochester, New York 14627, USA}
\author{M.~Artuso}
\author{S.~Blusk}
\author{S.~Khalil}
\author{R.~Mountain}
\author{K.~Randrianarivony}
\author{T.~Skwarnicki}
\author{J.~C.~Wang}
\author{L.~M.~Zhang}
\affiliation{Syracuse University, Syracuse, New York 13244, USA}
\collaboration{CLEO Collaboration}
\noaffiliation

\date{September 29, 2009}

\begin{abstract} 
We report evidence for the ground state of bottomonium, $\eta_b(1S)$, in the radiative decay 
$\Upsilon(3S)\to\gamma\eta_b$ in $e^+e^-$ annihilation data taken with the CLEO~III detector.  Using 6~million $\Upsilon(3S)$ decays, and assuming $\Gamma(\eta_b) = 10\; \mathrm{MeV}\!/c^2$, we obtain $\mathcal{B}(\Upsilon(3S)\to\gamma\eta_b) =(7.1\pm 1.8\pm 1.3)\times10^{-4}$, 
where the first error is statistical and the second is systematic.  The statistical significance is $\sim\!4\,\sigma$.  The mass is determined to be $M(\eta_b) = 9391.8 \pm 6.6 \pm 2.0\;\mathrm{MeV}\!/c^2$, which corresponds to the hyperfine splitting $\Delta M_{hf}(1S)_b = 68.5 \pm 6.6 \pm 2.0\;\mathrm{MeV}\!/c^2$.  Using 9~million $\utwos$ decays, we place an upper limit on the corresponding $\utwos$ decay, $\butwos < 8.4 \times 10^{-4}$ at $90\%$ confidence level.
\end{abstract}

\maketitle

The spectroscopy of the $b\bar{b}$ bottomonium states provides valuable insight into Quantum Chromodynamics (QCD) since relativistic and higher-order $\alpha_s$ corrections are less important for $b\bar{b}$ than any other $q\bar{q}$ system. Experimental measurements of the spectroscopic properties of the bottomonium states can therefore be compared with greater confidence with the predictions of perturbative QCD, as well as with lattice calculations. The hyperfine mass splitting of the singlet-triplet states is of particular interest since it probes the spin-dependent properties of the $q\bar{q}$ system.

The triplet $S~$state $(1^3S_1)$ of $b\bar{b}$ bottomonium, $\uones$, was discovered thirty years ago, but the identification of its partner, the singlet $S~$state $(1^1S_0)$, $\etabones$ (henceforth $\etab$), has eluded numerous searches, including those by CUSB~\cite{etabcusb}, ALEPH~\cite{etabaleph}, DELPHI~\cite{etabdelphi}, and CLEO~\cite{cleoiii_incl_rad}.  As a result, the $1S$ hyperfine splitting, which is well-determined in the charmonium system, remained unknown in the bottomonium system.  Recently, using their data sample of 109~million $\uthrees$ events, the BaBar collaboration reported~\cite{babar_etab,babar_etab2s} the observation of the $\etab$ with a statistical significance of more than $10\sigma$ (standard deviations) in the inclusive photon spectrum of $\uthrees$ 
with the observed photon energy 
$E_\gamma(\uthrees\to\gamma\etab) = 921.2^{+2.1}_{-2.8}\pm2.4\;\,\mev$, where the first error is statistical and the second is systematic.  This gave $M(\etab)=9388.9^{+3.1}_{-2.3}\pm2.7\;\,\mevcc$ and a bottomonium hyperfine splitting, $\Delta M_{hf}(1S)_b\equiv M(\Upsilon(1S))-M(\etab)=71.4^{+3.1}_{-2.3}\pm2.7\;\,\mevcc$.  BaBar's measured branching fraction was $\buthrees = (4.8\pm 0.5\pm 0.6)\times10^{-4}$.  Corroboration of the BaBar finding with an independent data set is essential. 

In this article we reexamine the CLEO data for the radiative decays $\Upsilon(3S,2S)\to\gamma\etab$.  
An earlier analysis of the same data resulted in upper limits of $\buthrees < 4.3\times10^{-4}$ and $\butwos < 5.1\times10^{-4}$ at $90\%$ confidence level~\cite{cleoiii_incl_rad}. However, the analysis had shortcomings which are rectified in this article.  The presence of the photon line corresponding to initial state radiation (ISR), $e^+e^-\to\gamma\Upsilon(1S)$, located between the $\chi_{bJ}(2P,1P)\to\gamma\uones$ region and the $\etab$ signal region, was not included in the fits to the inclusive photon spectrum, an omission which biased the result toward small branching fractions. The assumption of $\Gamma(\etab)=0$~MeV had a similar effect. Moreover, the analysis did not employ an important background-suppression variable, the angle between the radiative photon and the thrust axis of the rest of the event, introduced by BaBar~\cite{babar_etab}. We improve upon the previous publication by exploiting a more complete understanding of the expected photon line shape over a broad energy range to more accurately represent the $\chi_{bJ}(2P,1P)\to\gamma\uones$, ISR, and $\etab$ (with non-zero width) signals in a fit. We also employ a broader range of binning, fit ranges, and background parameterizations in order to avoid bias in any of these choices.

The CLEO III detector, which has been described elsewhere~\cite{cleodetector}, contains a CsI electromagnetic calorimeter, an inner silicon vertex detector, a central drift chamber, and a ring-imaging Cherenkov (RICH) detector, inside a superconducting solenoid magnet with a 1.5 T magnetic field. The detector has a total acceptance of 93$\%$ of $4\pi$.  The photon energy resolution in the central ($83\%$ of $4\pi$) part of the calorimeter is about $2\%$ at $E_{\gamma}=1~\gev$ and about $5\%$ at $100\;\,\mev$.  The charged particle momentum resolution is about 0.6$\%$ at $1~\gevc$.

The CLEO datasets correspond to $(5.88\pm 0.12)\times10^6~\Upsilon(3S)$ and $(9.32\pm 0.19)\times10^6~\Upsilon(2S)$ decays.  Our event selection for the inclusive photon spectra is identical to that reported in Ref.~\cite{cleoiii_incl_rad}.  Events are required to have one or more photons, and three or more charged tracks.  Photons with $E_\gamma\ge 20\;\,\mev$ are accepted in the ``good barrel'' region of the calorimeter with $|\cos\theta|<0.81$ (where $\theta$ is the polar angle with respect to the incoming positron direction), and are required to have a transverse spread which is consistent with that of an electromagnetic shower.  Photons from $\pi^0$ decays are suppressed by vetoing any photon candidates that, when paired with another photon candidate in the good barrel or ``good endcap'' ($0.85 < |\cos\theta| < 0.93$) regions, have a mass within $2.5\sigma$ of the known $\pi^0$ mass and $\cos\theta_{\gamma\gamma}>0.7$, where $\theta_{\gamma\gamma}$ is the opening angle of the photon candidates in the lab frame.

We first consider the analysis of the inclusive photon spectrum from $\uthrees$ decays.   The analysis of $\utwos$ decays follows
a similar path.  In the region $500 < E_\gamma < 1200\;\,\mev$, the spectrum consists of a peak centered around $E_\gamma\approx770\;\,\mev$
due to the three unresolved transitions, $\chi_{bJ}(2P)\to\gamma\uones$, $J=0,1,2$ on top of a smooth background that falls sharply
with energy. The peaks due to ISR and $\etab$, which are more than an order of magnitude weaker than those from $\chi_{bJ}(2P)$, are
expected in the high energy tail region of the $\chi_{bJ}(2P)$ peak.  Hence, sensitivity to the possible presence of an $\etab$
signal depends critically upon properly representing the shape of the $\chi_{bJ}(2P)$ peaks as well as suppressing the underlying
smooth background (as already achieved in part by the $\pi^0$ veto).  As demonstrated by the BaBar analysis~\cite{babar_etab},
additional suppression can be achieved by recognizing that $\etab$ signal photons are largely uncorrelated in direction with the
rest of the event, whereas background photons from the continuum tend to follow the leading particles of the underlying event. This
effect is more pronounced for $\uthrees\to\gamma\etab$ decays than for $\utwos\to\gamma\etab$, but the effect is nevertheless useful
for background suppression in both processes. The thrust angle ($\theta_T$) is utilized to exploit these correlations;  $\theta_T$
is determined for each event as the angle between the momentum vector of the signal photon and the thrust vector~\cite{thrust}
calculated using all \textit{other} final state photons and charged particles boosted into the rest frame of the $\etab$ candidate
(defined by the signal photon).  As shown in Fig.~1(a), the thrust angle distribution for the data events is peaked near
$|\cos\theta_T|=1$, whereas the thrust angle for the $\etab$ signal events from Monte Carlo (MC) simulations is distributed uniformly.  As a
result, the sensitivity to a possible $\etab$ signal in the presence of background varies greatly with $|\cos\theta_T|$, and it can
be maximized by taking advantage of the $|\cos\theta_T|$ distribution.

\begin{figure}[htb]
\begin{center}
\includegraphics[width=\figwidth]{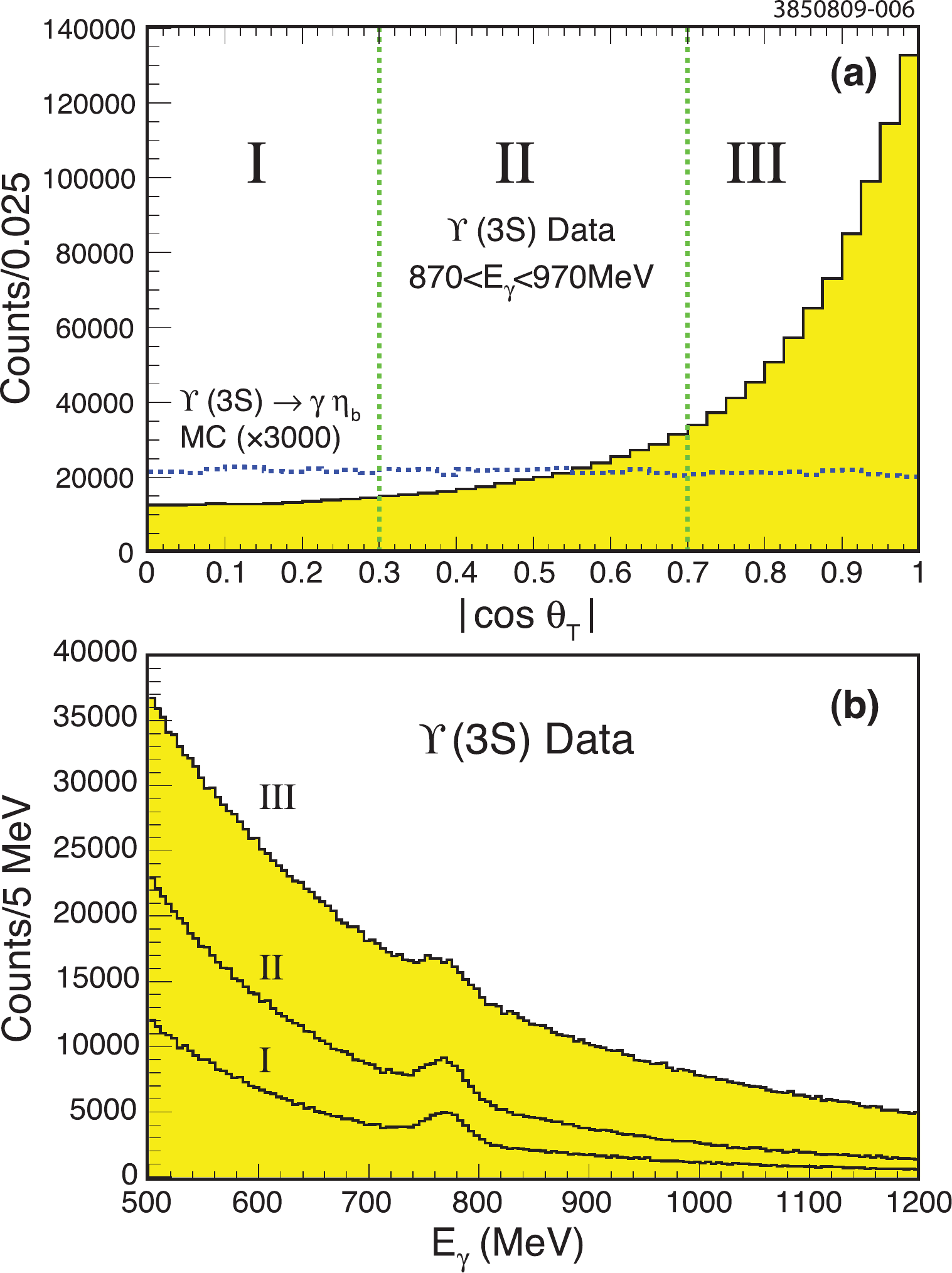}
\end{center}
\caption{(a) Distribution of $|\cos\theta_T|$ for $\etab$ signal MC events (dotted) and background dominated $\uthrees$ data (shaded) in three regions, I, II, and III defined in the text. The histogram of the Monte Carlo simulation of the $\uthrees\to\gamma\etab$ signal has been multiplied by a factor of 3000 to make it visible. (b) The $E_{\gamma}$ distribution from $\uthrees$ data in the three regions of $|\cos\theta_T|$.  Only the $\chi_{bJ}(2P)\to\gamma\uones$ lines at around $770\;\,\mev$ are visible above the background.}
\end{figure}

We utilize the $|\cos\theta_T|$ distribution, but in a manner quite different from that used by BaBar~\cite{babar_etab}.  Instead of simply rejecting all events with large values of $|\cos\theta_T|$, we increase the sensitivity to $\etab$ by forming three separate photon energy spectra, one each for the $|\cos\theta_T|$ regions $(0.0, 0.3)~\mathrm{(I)}$, $(0.3,0.7)~\mathrm{(II)}$, and $(0.7,1.0)~\mathrm{(III)}$, and performing a simultaneous joint fit to all three distributions.  The signal-to-background ratio improves from region~III to region~II and from region~II to region~I, but all regions contribute to the sensitivity.   Monte Carlo simulations show that, for a data sample of our size and a $\buthrees$ whose value is assumed to be what is measured below, the three-region joint fit procedure leads to an average increase in the statistical significance
\footnote{We compute the statistical significance of the fit using the conventional likelihood expression  
$\sqrt{2\ln(\calL_{sig}/\calL_0)}$, where $\calL_{sig}$ is the likelihood of the fit with a signal and $\calL_0$ is the likelihood of the fit with the signal constrained to zero.}
of an $\etab$ signal of $0.6\sigma$ over only accepting events with $|\cos\theta_T|<0.7$, albeit with a large r.m.s.\ spread of $0.7\sigma$ among MC trials. An average gain in significance over using no information about the thrust axis is $1.7\sigma$ with an r.m.s.\ spread of $1.6\sigma$.
Most of the $0.6\sigma$ increase in sensitivity from the joint fit comes from splitting the 
$|\cos\theta_T| < 0.7$ region into two bins, which exploits the smaller background relative to 
expected signal in the $|\cos\theta_T| < 0.3$ bin compared to the $0.3 < |\cos\theta_T| < 0.7$ bin.  
On the average, inclusion of the $|\cos\theta_T| > 0.7$ region by itself improves the result by $0.2\sigma$.  

The photon peaks have shapes which are parameterized by convolving a relativistic Breit-Wigner resonance function with a Crystal Ball~(CB) calorimeter response function~\cite{cb}, which consists of a Gaussian part with width $\sigma$ (the energy resolution) smoothly joined to a low-side power-law tail described by two additional shape parameters.  The energy resolution and CB shape parameters were determined with two complementary methods. In Method~A, we utilized isolated photons in $e^+e^-\to e^+e^-\gamma$ data events with photon energies near $E_{\rm true}=750\;\,\mev$, where $E_{\rm true}$ is the photon energy expected from using only the measured angles of the $e^\pm$ and $\gamma$. We then extracted an inherent line shape by deconvolving the spread in $E_{\rm true}$ (obtained from simulated events) from the observed $E_\gamma/E_{\rm true}$.  In Method~B, we compared exclusive $\Upsilon(3S)\to\gamma\chi_{b1}(2P)$, $\chi_{b1}(2P)\to\gamma\Upsilon(1S)$, $\Upsilon(1S)\to\ell^+\ell^-$ ($\ell^\pm\equiv e^\pm$ or $\mu^\pm$) in data and MC simulation to determine the shape of the $\Upsilon(3S)\to\gamma\chi_{b1}(2P)$ photon line.  The data distribution was used to determine the Gaussian part of the shape and the MC simulations were used to determine the two tail parameters after tuning the MC parameters to match the Gaussian part observed in the data.  Methods A and B lead to consistent energy resolutions and CB shape parameters, resulting in a line shape that is significantly different from that used in the original CLEO analysis. 
While the tail parameters of the peak shapes are fixed to be the same for all three relevant photon energies ($\chi_{bJ}(2P)$, ISR, and $\etab$), the Gaussian widths for the three are different.
The fitted Gaussian width for the overlapping $\chi_{bJ}(2P)$ peaks near 770~MeV in the inclusive spectrum is 
$\sigma(770\;\,\mev)=16.7\pm1.0\;\,\mev$.  
The variation of the photon resolution width with energy was determined
from MC simulations made for a wide range of photon energies. Its
parametrization was used to obtain the extrapolated values, 
$\sigma(859\;\,\mev)=17.4\pm1.0\;\,\mev$, and 
$\sigma(920\;\,\mev)=18.3\pm1.1\;\,\mev$, for the ISR and $\etab$ peaks, respectively.

The expected intensity of the ISR peak was obtained by extrapolating its yield observed in CLEO data taken on the $\ufours$ resonance.  The expected yield $N(\mathrm{ISR})=1726\pm131$, photon energy $E_\gamma(\mathrm{ISR}) = 859\;\,\mev$, and energy resolution $\sigma(\mathrm{ISR})=17.4\;\,\mev$ are fixed in all fits of the inclusive spectra.

The prominent peaks in the inclusive spectra shown in Fig. 1(b) are composites of the three 
$\chi_{bJ}(2P)\to\gamma\uones$ peaks for $J=0,1,2$.  We fix the relative strengths of these three lines to the ratios determined from other measurements~\cite{pdg} and float only the overall amplitude.  We also fix the spin-orbit splitting of these lines to the values measured in Ref.~\cite{cleoiii_incl_rad}, but we float the absolute energy scale. The latter provides a useful check on our uncertainty in the absolute energy calibration. The CB~line shape parameters are fixed as discussed previously, while the effective energy resolution, which includes Doppler smearing, is allowed to float.

The efficiencies for $\chi_{bJ}(2P)$, ISR, and $\etab$ in our event selections are obtained by Monte Carlo simulations with the $1+\alpha\cos^2\theta$ angular distributions expected for E1 and M1 transitions with appropriate values of $\alpha$ for $\chi_{b1}(2P)$ and $\chi_{b2}(2P)$, and $\alpha=1$ for $\chi_{b0}(2P)$ and $\etab$.  Separate calculations were done for the three $|\cos\theta_T|$ bins, and it was found that efficiencies are approximately constant in $|\cos\theta_T|$.  The summed efficiencies for $\etab$ and ISR are $(54.2\pm3.8)\%$ and $(6.9\pm0.1)$\%, respectively.

As discussed previously, we perform a joint fit of the data in three $|\cos\theta_T|$ bins.  All fitting parameters (apart from those in the background function described below) are constrained to be the same in the three $|\cos\theta_T|$ bins.  That is, the yields for the $\chi_{bJ}(2P)$,
ISR, and $\etab$ photon peaks in each of the three $|\cos\theta_T|$ bins were constrained to be
proportional to the ratios $\Delta|\cos\theta_T|_i/\epsilon_i$ where $\epsilon_i$ is the signal 
efficiency for bin $i$.

The smooth background was fitted with exponential polynomials,
\begin{equation}
\frac{dN}{dE_\gamma} = \exp\left( \sum_{i=0}^{i=n} a_i E_\gamma^i \right).
\end{equation}
As the only experimental handle on these backgrounds is the inclusive spectrum itself, we explored uncertainties in their determination by varying binning types (both linear and logarithmic binning were used), the order  
of the polynomial ($n$ was varied from 2 to 4 in each thrust bin independently) and the fit range (six different ranges were tried extending down to $500\;\,\mev$ and up to $1340\;\,\mev$). Results for the $\etab$ (mass, significance, and branching fraction) were then averaged through all fits with confidence level (CL) above $10\%$.  The r.m.s.\ spread among the fit variations was taken as a measure of the systematic uncertainty in the background determination. Averaged through all successful fits, the maximum likelihood significance of the $\etab$ signal is $4.1\sigma$ with a r.m.s.\ spread of 0.4.  A representative fit, whose parameters are close to the average values for the ensemble of accepted fits, is chosen as nominal.  This fit (shown in Fig.~2) has $N(\gamma\etab)=2311\pm546$~counts and gives 
$\bgammaetab \equiv \buthrees = (7.1\pm1.8)\times10^{-4}$ and 
$\egammaetab \equiv E_\gamma(\uthrees\to\gamma\etab)= 918.6\pm6.0\;\,\mev$, 
with a CL of 18.5\%.

\begin{figure}[htb]
\begin{center}
\includegraphics[width=\figwidth]{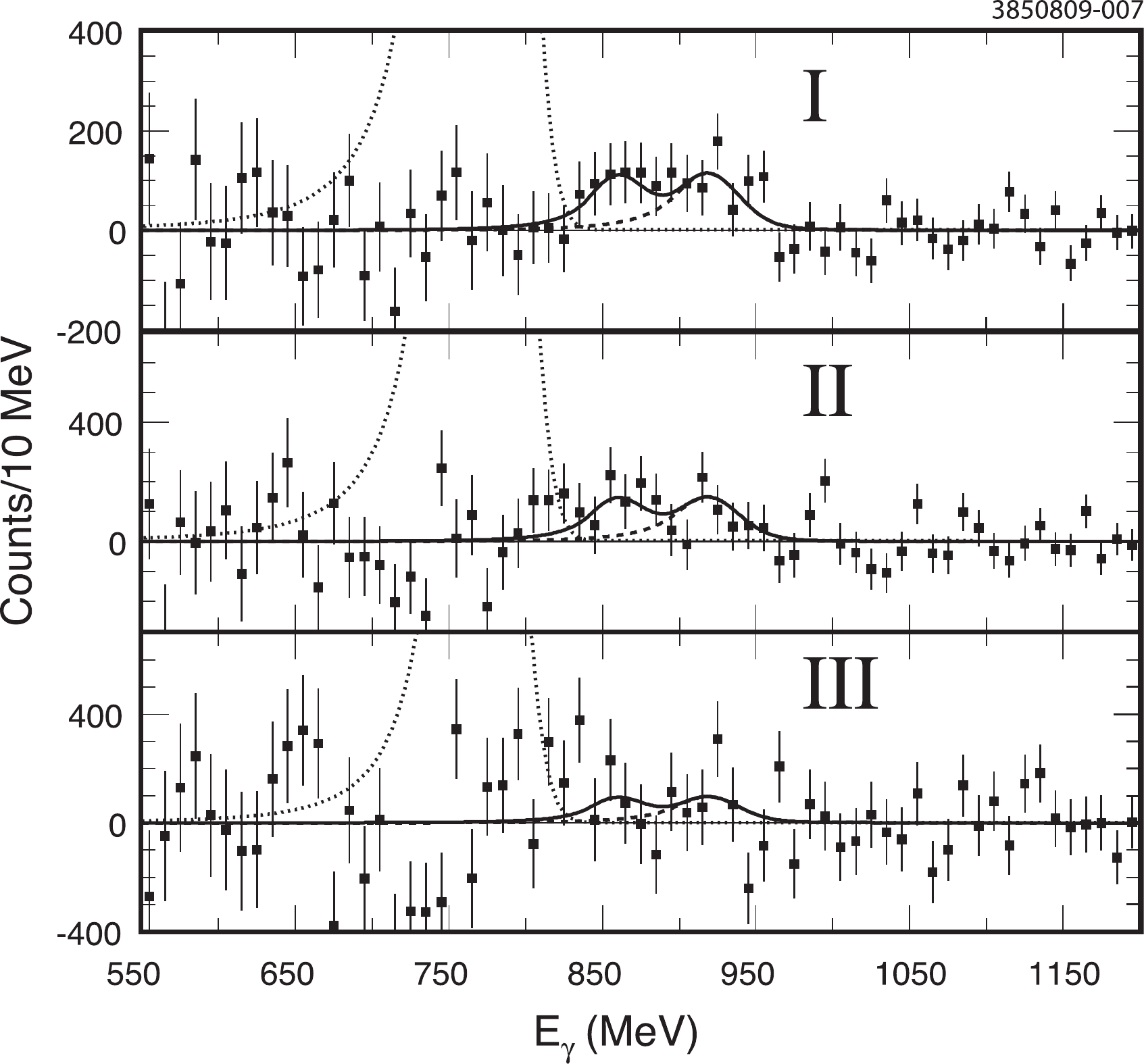}
\end{center}
\caption{Background and $\chi_{bJ}(2P)$ subtracted distributions of $E_\gamma$ from $\uthrees$ decays in three $|\cos\theta_T|$ regions, I, II, and III defined in the text.  The curves are the results of the joint fit, with a CL of 18.5\%.  The $\chi_{bJ}(2P)$ peaks are indicated by the dotted lines and the $\etab$ signals by the dashed lines, which join the solid line.}
\end{figure}

\begin{table}[htb]
\caption{Summary of estimated systematic uncertainties and their sums in quadrature for the 
$\uthrees\to\gamma\etab$ analysis.  The item labeled Background refers to variations of the background function parameters, the fit range, and linear versus logarithmic $E_{\gamma}$ binning.}
\begin{center}
\begin{tabular}{lcc}
\hline \hline
 & \multicolumn{2}{c}{Uncertainty in} \\
Source & ${\egammaetab}$ (MeV) & ~${\bgammaetab}$ (\%)~ \\
\hline
Background   & $\pm1.0$ & $\pm10$~ \\
Photon Energy Calibration & $\pm1.2$ & --- \\
Photon Energy Resolution  & $\pm0.3$ & $\pm2$ \\
CB and $\chi_{bJ}(2P)$ Parameters   &    $\pm0.7$ & $\pm8$ \\
ISR Yield & $\pm0.4$ & $\pm3$ \\
Photon Reconstruction & --- & $\pm2$ \\
$N(\Upsilon(3S))$ & --- & $\pm2$ \\
MC Efficiency  & --- & $\pm7$ \\
$\eta_b$ Width & $\pm 0.6$ & $\pm 9$ \\ 
\hline
Quadrature sums & $\pm1.9$ & $\pm 18$~~ \\ 
\hline \hline
\end{tabular}
\end{center}
\end{table}

The systematic uncertainties in our results are obtained as follows and are summarized in Table~I.  We assign the r.m.s.\ variations in the results obtained for all the accepted fits, 
$\pm1.0\;\,\mev$ in  $\egammaetab$, and $\pm10\%$ in $\bgammaetab$ as systematic uncertainties 
due to background shape, binning, and range variations.  The changes in our results are negligible when
we alter the lower CL limit for acceptable fits from 10\% to either 5\% or 15\%.
We vary the photon energy resolution, the Crystal Ball shape parameters, and the $\chi_{bJ}(2P)$ parameters within their errors and assign the resulting variations in $\egammaetab$ and $\bgammaetab$ as systematic uncertainties.

The fitted $\chi_{bJ}(2P)$ centroid energy in our data is $769.9\pm0.2\;\,\mev$, while the expected energy is $769.6^{+0.7}_{-1.0}\;\,\mev$.  The $0.3\;\,\mev$ deviation of our measured value suggests that our photon energy calibration has a maximum possible uncertainty of $^{+0.9}_{-1.2}\;\,\mev$.  This is consistent with our measurement of ISR photon energies from $\Upsilon(4S)$ and below $\Upsilon(4S)$ data, which agree with the expected energies within $\pm0.3\;\,\mev$.  Based on these considerations we conservatively assign the systematic uncertainty due to photon energy calibration as $\pm1.2\;\,\mev$. We obtained the value of 
$\bgammaetab$ by assuming $\Gamma(\etab)=10\;\,\mevcc$.  We find that $\bgammaetab$ depends linearly 
on the assumed value of $\Gamma(\etab)$ in $\;\,\mevcc$, as 
$\bgammaetab = [5.8+0.13\, \Gamma(\etab)]\times10^{-4}$.  
Varying $\Gamma(\etab)$ from $5$ to $15\;\,\mevcc$, a range that includes 
nearly all theoretical expectations~\cite{theorywidth}, 
the branching fraction changes by $\pm 0.65\times 10^{-4}$ or $\pm 9$\%.
This uncertainty in the $\etab$ width also contributes $\pm 0.6$~MeV to the uncertainty in 
$\egammaetab$.
Other systematic uncertainties are due to the Monte Carlo efficiency calculation and the 
number of $\Upsilon(3S)$ events.

In fitting the $\gamma\etab$ peaks, we do not include the   
factor~\cite{egamma-depend} $(\Egamma/\Egammaz)^3\, [1 + \alpha (\Egamma/\Egammaz)^2]^2$ 
expected in the decay width for the hindered M1 transition $\uthrees\to\gamma\etab$.  
($\Egammaz$ is the photon energy for the central value of the $\etab$ mass.)  
While theoretical estimates~\cite{egamma-depend} of alpha vary, $\alpha = 1$ 
leads to a distortion of the $\etab$ peak shape and a consequent 
reduction of $\egammaetab$ by approximately 3~MeV. Since our data sample is not 
large enough to determine $\alpha$, in the absence of firm theoretical predictions 
we do not include this effect as a bias or as a term in our systematic error.

Our final results are:  $\egammaetab = 918.6 \pm 6.0 \pm 1.9\;\,\mev$ and
$\bgammaetab = (7.1 \pm 1.8 \pm 1.3)\times 10^{-4}$, where the first errors are statistical 
and the second errors are systematic.
Our result for $\egammaetab$ corresponds to $M(\etab)=9391.8\pm6.6\pm2.0\;\,\mevcc$ and 
$\Delta M_{hf}(1S)_b=68.5\pm6.6\pm2.0\;\,\mevcc$.  This is consistent with lattice QCD 
predictions that employ dynamical quarks and include both continuum and chiral 
extrapolations~\cite{lqcd-etab}. 
Our results for both $\Delta M_{hf}(1S)_b$ and $\bgammaetab$ are
also well within the wide range of pQCD based theoretical predictions~\cite{theorygr}.  
Both measurements are in good agreement with the BaBar measurements~\cite{babar_etab,babar_etab2s}.

\begin{figure}[htb]
\begin{center}
\includegraphics[width=\figwidth]{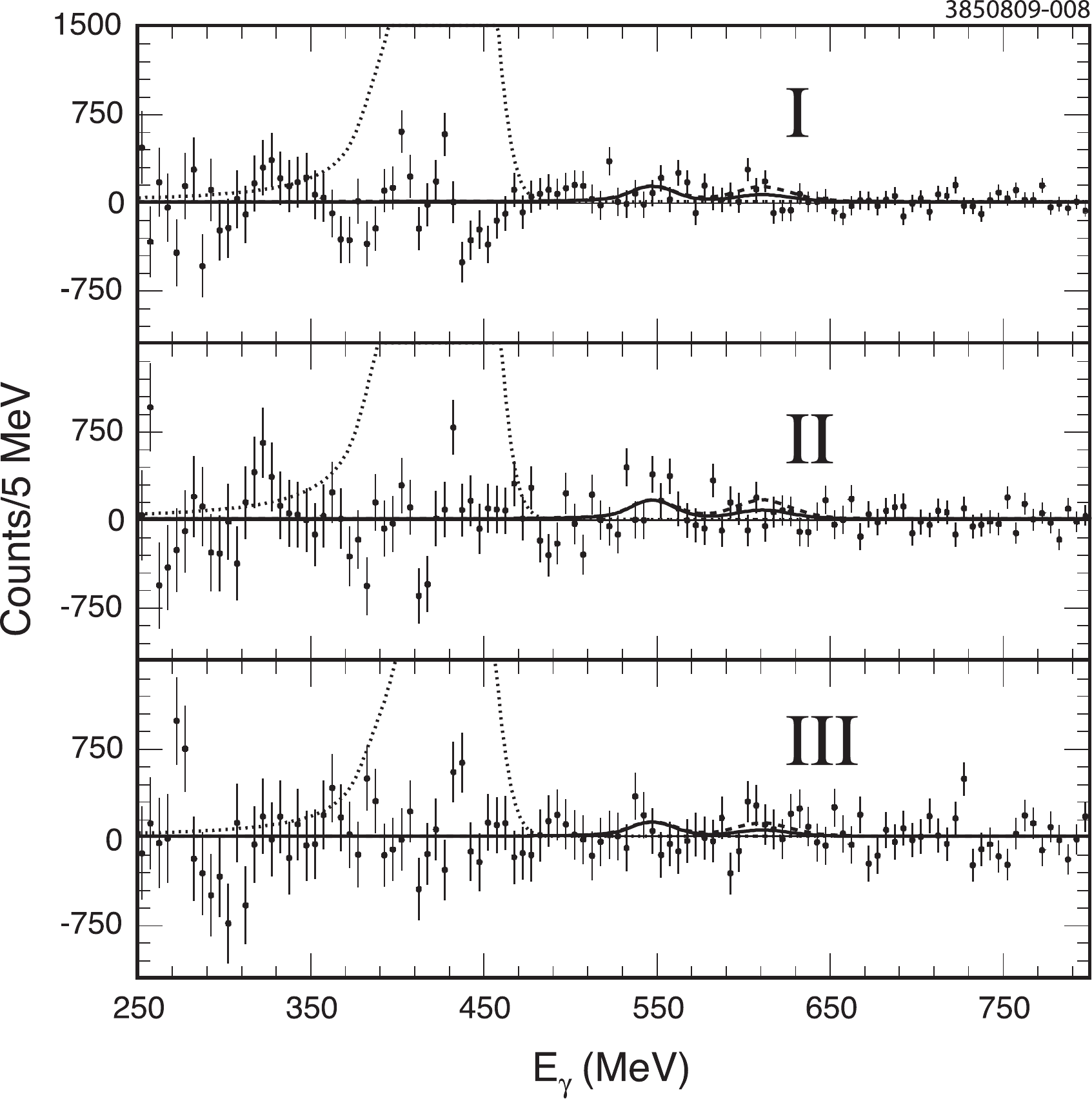}
\end{center}
\caption{Background and $\chi_{bJ}(1P)$ subtracted distributions of $E_\gamma$ from $\utwos$ decays in three $|\cos\theta_T|$ regions, I, II, and III defined in the text.  The curves are the joint fit results.  The $\chi_{bJ}(1P)$ peaks are indicated by the dotted lines and the 90\% $\etab$ upper limits by the dashed lines.}
\end{figure}

We also analyzed our data set containing $(9.32\pm0.19)\times10^6~\Upsilon(2S)$ events for $\Upsilon(2S)\to\gamma\etab$ using the same event selection and joint fit analysis procedure as described above for $\Upsilon(3S)\to\gamma\etab$.  One difference is that we chose to represent the $\utwos\to\pi^0\pi^0\uones$ background component explicitly in the fit since it introduces a kink in the spectrum not far from the signal region.  The shape of this background was taken from Monte Carlo simulations. Its normalization was fixed to the PDG value of the branching fraction.  Unlike in the $\utwos$ analysis, the addition of the explicit $\uthrees\to\pi^0\pi^0\uones$ background component to the $\uthrees$ fits had a negligible effect on the results.  In the expected signal region for $\utwos$ radiative decay, 
$200 < E_\gamma < 800\;\,\mev$, the background is an order of magnitude larger than in the $\uthrees$ signal region, and in none of the $\utwos$ $|\cos\theta_T|$ regions could the $\etab$ be identified.  In the joint fit analysis (shown in Fig.~3), fixing $E_\gamma(\utwos\to\gamma\etab) = 611\;\,\mev$, corresponding to $\etab$ mass determined in $\uthrees$ decay, leads to $\butwos = (3.9\pm2.7\pm2.3)\times10^{-4}$, or an upper limit of $\butwos <8.4 \times10^{-4}$ at 90\% confidence level.  This is consistent with the BaBar $\utwos$ result~\cite{babar_etab2s}, $\butwos)=(3.9^{+1.1}_{-1.0}\pm0.9)\times10^{-4}$.

We gratefully acknowledge the effort of the CESR staff in providing us  
with excellent luminosity and running conditions. This work was  
supported by the A.P.~Sloan Foundation, the National Science  
Foundation, the U.S. Department of Energy, the Natural Sciences and  
Engineering Research Council of Canada, and the U.K. Science and  
Technology Facilities Council.

\end{document}